\documentstyle[aps,prd,eqsecnum,epsf]{revtex}

\newcommand{\DS}{{\Delta S}}
\newcommand{\DSBH}{{\DS_{BH}}}
\newcommand{\DSM}{{\DS_M}}
\newcommand{\DEBH}{{\Delta E_{BH}}}
\newcommand{\DEM}{{\Delta E_M}}
\newcommand{\TBH}{T_{BH}}
\newcommand{\bBH}{\beta_{BH}}
\newcommand{\Ip}{{\cal I}^+}
\newcommand{\Imm}{{\cal I}^-}
\newcommand{\Hp}{{\cal H}^+}
\newcommand{\Hm}{{\cal H}^-}
\newcommand{\heps}{\hat{\epsilon}}
\newcommand{\hrho}{\hat{\rho}}
\newcommand{\hOmega}{\hat{\Omega}}

\newcommand{\hP}{\hat{P}}
\newcommand{\hU}{\hat{U}}
\newcommand{\hV}{\hat{V}}
\def\rslash{\partial\kern-0.026em\raise0.17ex\llap{/}%
\kern0.026em\relax}
\def\Dslash{D\kern-0.15em\raise0.17ex\llap{/}\kern0.15em\relax}
\def\sqr#1#2{{\vcenter{\hrule height.#2pt
      \hbox{\vrule width.#2pt height#1pt \kern#1pt
          \vrule width.#2pt}
      \hrule height.#2pt}}}

\begin{document}
\title{Does the generalized second law hold in the form of 
	time derivative expression?}

\author{T. Shimomura, T. Okamura, T. Mishima${}^{\dag}$ and H. Ishihara} 

\address{
Department of Physics, Tokyo Institute of Technology,
Oh-Okayama, Meguro, Tokyo 152-0033, Japan\\
${}^{\dag}$Laboratories of Physics, College of Science and Technology, 
Nihon University, Narashinodai, Funabashi, Chiba 274-0063, Japan}

\date{\today}

\maketitle

\begin{abstract}
\noindent
We investigate whether the generalized second law is valid, 
using two dimensional black hole spacetime, irrespective of models.
A time derivative form of the generalized second law
is formulated and it is shown that the law might become invalid.
The way to resolve this difficulty is also presented
and discussed.
\end{abstract}


\section{Introduction}\label{sec:ointro}


One of the most interesting developments in black hole physics is 
a discovery of the analogy between certain laws of black hole mechanics
and the ordinary laws of thermodynamics
\cite{BCH}. 
According to this analogy, Bekenstein
\cite{BK} 
introduced the concept of the black hole entropy as a quantity 
proportional to the surface area of the black hole 
(the proportionality coefficient was fixed by Hawking's discovery
of black hole radiance
\cite{HW}
later)
and conjectured that the total entropy never decrease in any process, 
where the total entropy is the sum of the black hole entropy and 
the ordinary thermodynamic entropy of the matter outside the black hole. 
This is known as the generalized second law(GSL) of thermodynamics
and it is important to check the validity of this conjecture
because the validity strongly supports that 
the ordinary laws of thermodynamics can
apply to a self-gravitating quantum system containing a black hole. 
Especially, it strongly suggests 
the notion that $A/4$ ($A$ is the surface area of the black hole)
truly represents the physical entropy of the black hole. 

In order to interpret $A/4$ as the black hole entropy, 
it would be necessary to derive $S_{BH}=A/4$ from a statistical
mechanical calculation by counting the number of internal states 
of the black hole.
The microscopic derivation of the black hole entropy along this line 
achieved some results in the recent progress in superstring theory
\cite{SUPER}.
However, general arguments for the validity of the second law of 
thermodynamics for ordinary systems are based on notions of 
the {\lq}{\lq}fraction of time''
a system spends in a given macroscopic state. 
Since the nature of time in general relativity is drastically different 
from that in nongravitational physics, it is not clear how the GSL will 
arise even if $A/4$ represents a measure of the number of internal states 
of the black hole. 
Therefore, it is important to examine 
the validity of the GSL by itself, 
in order to understand the connection 
between quantum theory, gravitation and thermodynamics further.

Historically, gedanken experiments have been done to test 
the validity of the GSL. The most famous one is that in which a box
filled with matter is lowered to near the black hole and then dropped in
\cite{BK}. 
Classically, a violation of the GSL can be achieved if one lowers the
box sufficiently close to the horizon. However, when the quantum effects 
are properly taken into account, it was shown by Unruh and Wald
\cite{UW} 
that the GSL always holds in this process.

On the other hand, there are some people who tried to prove the GSL 
under several assumptions for more general situations.
Frolov and Page
\cite{FP}
proved the GSL for an eternal black hole by assuming that 
(i) the process in the investigation is quasistationary
which means that the change in the black hole geometry are sufficiently 
small compared with the corresponding background quantities,
(ii) the state of matter fields on the past horizon 
$\Hm$ is a thermal state with the Hawking temperature,
(iii) initial set of radiation modes on the past horizon
$\Hm$ and that on the past null infinity $\Imm$ 
are quantum mechanically uncorrelated,
and (iv) the Hilbert space and the Hamiltonian of modes at 
$\Hp$ are identical to those of modes at $\Hm$.

But these assumptions are questionable
for the black hole formed by a gravitational collapse.
That is, the assumptions (iii) and (iv) break down due to the correlation 
between modes at $\Hm$ and modes at $\Imm$
located after the horizon formation
and the violation of time reversal symmetry, respectively.
So we think that their proofs should be improved to realistic 
black holes formed by gravitational collapse.

The GSL for the black hole formed by gravitational collapse 
was studied by Sorkin
\cite{SK} 
and Mukohyama
\cite{MK},
making use of the nondecreasing function in a Markov process. 
The proof finally come to showing that the matter fields 
in the black hole background have a stationary canonical distribution 
with its temperature equal to that of the black hole and 
the canonical partition function remains a constant.

But there are several problems in their proofs. 
Mukohyama showed that the canonical distribution with temperature
equal to the black hole is stationary by calculating the transition
matrix between states at the future null infinity $\Ip$ and 
states at the portion of the past null infinity $\Imm$ 
after the formation of the event horizon $\Hp$.
But, in collapsing cases,
the assumption (i) can not be justified in general.
By contrast, 
since Sorkin argued any process occurring 
between two adjacent time slices, the assumption (i) is valid. 
He concluded that the canonical distribution of matter fields 
are stationary because the Hamiltonian does not change
between the two adjacent time slices,
thanks to the time translation invariance in the background.
Althougth he assumed implicitly the existence of 
the Killing time slices that do not 
go through the bifurcate point, and derived his result, 
there would not exist such Killing time slices that he had taken. 
If we take the Killing time slices, 
there is no energy flux across the event horizon.
It means that we cannot see evaporating black hole 
by the Killing time slices.

In order to satisfy the assumption (i),
we consider the infinitesimal time development of total entropy
in two dimensional theories of gravity.
Although we think two dimensional spacetime,
it is worth to investigate the GSL in two dimensional black hole spacetime 
if there also exists the same black hole physics 
as those in four dimensional one
(causal structure, Hawking radiation and so on).
Because, in this case, we can expect that the essential point of the four
dimensional physics would not be lost.

Using the Russo-Susskind-Thorlacius (RST) model
\cite{RST}, 
Fiola, et.al. discussed the infinitesimal time development 
of total entropy and showed that the GSL in the model is valid under 
suitable conditions.
Although their investigation is beyond the quasistationary approximation
and takes account of quantum-mechanical back-reaction effects,
their argument is restricted to the very special (RST) model
and it is too hasty in concluding that the GSL generally holds even 
in two dimensional spacetime. 
Because if {\lq}{\lq}black hole entropy'' truly represents the physical 
entropy of a black hole,
it would be necessary to confirm the validity of the GSL for the more 
general models which possess the black hole mechanics.

The purpose of this paper is to investigate the GSL in any two dimensional
black hole spacetime with the first law of black hole mechanics, 
irrespective of models.
In fact, the existence of the first law is guaranteed 
for the wide class of gravitational theories by using the Noether charge method
\cite{WD}.

First, we write the change in total entropy
between two adjacent time slices in terms of quantities of matter fields,
using the assumption (i) and the first law of black hole mechanics. 
Thus, our task is to calculate the energy-momentum tensor 
and the entanglement entropy of matter fields.
These are obtained easily for conformal fields
in two dimensional spacetime.
After these calculations,
we will demonstrate that 
the GSL does not always hold for conformal vacuum states 
in a two dimensional black hole for two reasons. 
The first is that the GSL is violated by the decrease of 
the entanglement entropy of the field associated with
the decrease of the size of the accessible region.
But it might be possible to subtract this term by some physical procedure 
and define the new entropy.
The second is that the GSL for the new entropy would be violated
for some class of the vacuum states.
It might suggest that even the GSL for the new entropy does not hold 
as long as there does not exist a physical reason
that exclude these vacuum states.
In this sense, there seem to exist two difficulties to rescue the GSL.

This paper is organized as follows.
In Sec.\ref{sec:review} we present two dimensional black holes
that we will consider in this paper 
and formulate a time derivative form of the GSL
Since we express the change in total entropy 
in terms of physical quantities of matters,
our task is to calculate a time evolution of matter fields 
in a fixed black hole background. 
In Sec.\ref{sec:evolI} we calculate the change in
energy of matter fields for general conformal vacuum states.
In Sec.\ref{sec:evolII} the entanglement entropy is obtained.
By way of illustration, these results are applied to the typical 
two vacuum states;
the Hartle-Hawking state and the Unruh one.
Then in Sec.\ref{sec:phys}, it is shown that the time derivative 
form of the GSL does not always hold for the general situations.
We will also give a physical interpretation of 
our result.
Sec.\ref{sec:summary} is devoted to summary and discussion about 
our results. In particular, we propose a new entropy and argue the validity
of the GSL for this quantity.


\section{Two dimensional black holes and the GSL}\label{sec:review}


\subsection{Two dimensional black holes}

Four dimensional gravitational theories have 
many degrees of freedom and inherent complexity.
So it would be useful to consider a toy model
in which greater analytic control is possible. 
In our analysis, we consider any two dimensional theories of gravity 
which satisfy the following two assumptions;
(1) the theory allows a stationary black hole solution,
and (2) there exist black hole physics similar to those in four dimensional
gravitational theories.

Since we want to examine the validity of the GSL in the two dimensional
eternal black hole background, we assume by assumption (1) that 
the spacetime possesses an event horizon and a timelike Killing vector.
Since we can always take spacelike hypersurface which is orthogonal
to the orbits of the isometry in two dimensional spacetime, 
this means that the theory has a static black hole solution as
\begin{eqnarray}
	ds^2&=&-\xi^2(r) dt^2+ {dr^2 \over \xi^2(r)}~
\label{eq:CGHSBH} \\
	&=&-\xi^2~dx^+ dx^-,
\end{eqnarray}
where $\xi^\mu=(\partial/\partial t)^\mu$ is the timelike Killing vector
which is normalized s.t. $\xi^2\rightarrow 1$ as $r\rightarrow\infty$, 
$x^\pm = t \pm r_*$ and $r_* =\int dr/{\xi}^2$, respectively.
The position of the horizon ${\cal H}$ is specified by $r$ with $\xi(r) = 0$ 
and the surface gravity of the black hole is given by
$\kappa = {\partial_{+} \ln{\xi^2}}{|}_{\cal H}$.

By assumption (2), we require 
that the black hole satisfies the first law of black hole mechanics. 
This is necessary to formulate the GSL in the form of the next subsection
and to keep the essential features of four dimensional black hole physics 
in our toy model. 

In fact, Wald
\cite{WD}
derived a first law of black hole mechanics 
for any diffeomorphism invariant gravitational theories
in any dimensions relied on the Noether charge associated
with the diffeomorphism invariance of the action
\footnote{
In the Euclidean method, we can get the Hawking temperature 
$T_{BH}=\kappa / 2\pi$
by requiring the nonsingularity of the Euclidean metric
\cite{GH}.
Thus, we regard the quantity $\kappa / 2\pi$
as the Hawking temperature in the Noether charge method.
}.
His technique is a quite general approach for a stationary black holes 
with a Killing horizon, and reproduces a known result
for Einstein gravity with ordinary matter actions.
Since gravitational theories are generally defined from a diffeomorphism 
invariant action, this assumption seems to hold naturally
for a very broad class of gravitational theories.

One evidence to justify these assumption is the existence of an 
interesting toy model which satisfy these assumptions. 
It is known that the CGHS model
\cite{CGHS}
has a static black hole solution
which evaporates by the Hawking effect, semiclassically. 
Moreover, the thermodynamical nature of this solution 
had been investigated by Frolov
\cite{FR}
and shown that black holes in the CGHS model also satisfy
the three laws (including the first law of black hole mechanics)
similar to {\lq}{\lq}standard" four dimensional black hole physics.

Therefore, it is quite reasonable to think that these two assumptions
hold for a wide class of gravitational theories.



\subsection{The GSL under the quasistationary approximation}


Before examining whether the GSL holds,
we formulate a precise statement of the GSL
in the quasistationary approximation.

As stated in the introduction, we foliate our black hole spacetime
by spacelike time slices that are across the event horizon 
and do not cross one another at the horizon.
We take two adjacent time slices among them 
to consider a quasistationary process 
(to justify the assumption (i) in the introduction)
and consider the change in total entropy between two time slices.

Under the situation satisfying the assumption (i), 
by making use of the first law of black hole mechanics 
($\DSBH=\DEBH/\TBH\equiv\bBH\DEBH$),
we can rewrite the change in total entropy 
in terms of quantities of matter fields alone,
\begin{eqnarray}
	\DS_{total} &=& \DSM + \DSBH
\label{eq:SECL} \\
	&=& \DSM + \bBH \DEBH
\label{eq:ECON} \\	
	&=& \DSM - \bBH \DEM~,
\end{eqnarray}
where we used the energy conservation law ($\DEBH=-\DEM$)
in the last line.

Thus, our task is to calculate 
the change in energy and entropy of matter fields 
between two adjacent time slices in the black hole background.
We will use the entanglement entropy of matter
fields outside the horizon as the quantity $S_M$.
These are obtained easily for massless conformal fields 
in two dimensional spacetime.

Further if we define the free energy of matter fields $F_{M}$ 
by $F_{M} \equiv E_M-\bBH^{-1} S_M$, 
we can write
\begin{equation}
	\DS_{total} = -\bBH \Delta F_{M}~,
\end{equation}
and to prove the GSL in the quasistationary approximation 
is equivalent to show that the free energy $F_{M}$ 
is a monotonically decreasing function of time. 
So we will be concerned with examining
the change in free energy of matter fields
immersed in the black hole background as a heat bath.


\section{The calculation of the energy-momentum tensor}

\label{sec:evolI}


For two dimensional conformal fields, 
we can obtain the energy-momentum tensor $T_{\mu\nu}$ 
by using its transformation law under conformal transformations
and the trace anomaly formula, which is for a scalar field,
\begin{equation}
	T={R \over 24\pi}~.
\end{equation}

When the spacetime metric in interest is given by
\begin{eqnarray}
	& &ds^2=-\hat{\Omega}^2~d\hat{U} d\hat{V},
\label{eq:COMH}
\end{eqnarray}

$T_{\mu\nu}$ for a conformal scalar field is written as
\cite{BD}
\begin{eqnarray}
	& &T_{\mu\nu}[g_{a b}]=T_{\mu\nu}[\eta_{a b}]
	+\theta_{\mu\nu}+{T[g_{a b}] \over 2}g_{\mu\nu},
\label{eq:TmunuORI} \\
	& &\theta_{\hat{U} \hat{U}}={-1 \over 12\pi}
	\hOmega \partial^2_{\hat{U}} {\hOmega}^{-1},
\nonumber \\
	& &\theta_{\hat{V} \hat{V}}={-1 \over 12\pi}
	\hat{\Omega} \partial^2_{\hat{V}} \hat{\Omega}^{-1},
\nonumber \\
	& &\theta_{\hat{U} \hat{V}}=\theta_{\hat{V} \hat{U}}=0,
\nonumber \\
\end{eqnarray}
where the null coordinate $(\hat{U},\hat{V})$ is given by 
$\hat{U} = \hat{U}(x^-), \hat{V} = \hat{V}(x^+)$,
and $g_{a b}=\hat{\Omega}^2\eta_{a b}$, respectively. 
Note that if we take the conformal vacuum, then the first term on
the R.H.S. of Eq.(\ref{eq:TmunuORI}) vanishes.

\begin{figure}
\begin{center}
\leavevmode
\epsfysize=7.0cm
\epsfbox{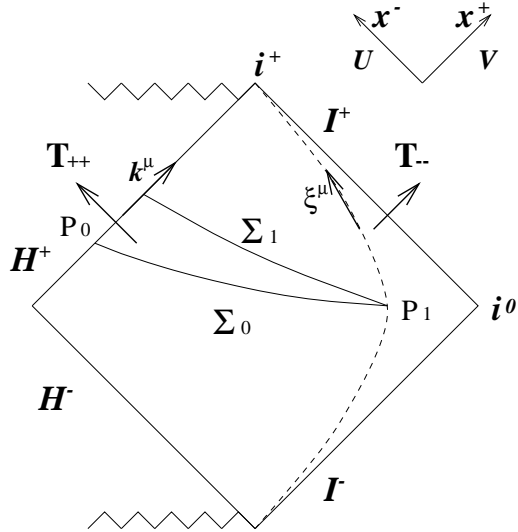}
\end{center}
\caption{This figure shows schematic explanation of situation considered here,
and the relation between the quantities that is necessary to calculate 
the change in energy between two spacelike time slices $\Sigma_{0}$ and 
$\Sigma_{1}$.}
\label{fig:coord}
\end{figure}

Now we apply this result to a two dimensional black hole in the previous
section.
We assume that we take the conformal vacuum associated with the $\hat{U}$
and $\hat{V}$.
Due to the existence of the timelike Killing vector
$\xi^\mu=(\partial/\partial t)^\mu$,
the quantity $E_{\hat{\Omega}}=-\int_\Sigma d\Sigma^\mu T_{\mu\nu} \xi^\nu$
is a function of the boundary of $\Sigma$.
We put the inner boundary $P_0=(x^+_0,x^-_0)$ of $\Sigma$ 
on the future horizon $\cal{H}^+$ 
and fix the outer boundary $P_1=(x^+_1,x^-_1)$ apart from the black hole.
And then, we consider the change in $E_{\hat{\Omega}}(x^+_0)$ by moving 
the inner boundary $P_0$ along $\Hp$.
This is given by
\begin{equation}
	{d E_{\hat{\Omega}}(x^+_0) \over d x^+_0} = - T_{++}|_{\Hp}~.
\end{equation}
These relations are sketched in
Fig.\ref{fig:coord}.

Then,  we can rewrite $T_{++}$ by using the relations Eq.(\ref{eq:COMH}) as
\begin{eqnarray}
	T_{++} &=& (\partial_{+}\hat{V})^2 T_{\hat{V}\hat{V}}
\\
	           &=& {-(\partial_{+}\hat{V})^2  \over 12\pi}
	           \hat{\Omega} \partial^2_{\hat{V}} \hat{\Omega}^{-1}
\\
	           &=&{1  \over 12\pi} \left[
	           {\partial^2_{+} \hat{\Omega} \over \hat{\Omega}} - 
	           {\partial_{+} {\xi}^2 \over {\xi}^2}
	           {\partial_{+} \hat{\Omega} \over \hat{\Omega}} \right].
\end{eqnarray}
and can express the change in $E_{\hat{\Omega}}(x^+_0)$ in terms of the 
conformal factor 
$\hat{\Omega}$;
\begin{eqnarray}
              {d E_{\hat{\Omega}}(x^+_0) \over d x^+} &=& - T_{++}|_{{\cal{H}}^{+}}
\\              
               &=&{-1  \over 12\pi} \left[
	           {\partial^2_{+} \hat{\Omega} \over \hat{\Omega}}\Big{|}_{\Hp} - \kappa
	           {\partial_{+} \hat{\Omega} \over \hat{\Omega}}\Big{|}_{\Hp} \right].
\label{eq:CHANGEE}
\end{eqnarray}

Next, we apply the above result to the typical two vacuum states;
the Hartle-Hawking one ($HH$) and the Unruh one ($U$).

For simplicity, we consider the case in which the horizons are
not degenerate. 
If there exist several horizons, we take the most outer one
and proceed to our arguments. 
In this case, we can rewrite the metric as follows; 
\begin{eqnarray}
	& &ds^2=-\Omega^2_{HH}~ dU dV 
	= -\Omega^2_U~ dU dx^+~,
\\	
	& &\Omega_{HH}^2={\xi^2 \over \kappa^2 |U V|}~,
\\
	& &\Omega_U^2\equiv \Omega^2_{HH} {dV \over dx^+}
	=\Omega^2_{HH} e^{\kappa x^+}~,
\end{eqnarray}
where the null coordinate $(U,V)$ is a Kruskal like one 
which is regular at the horizon ${\cal H}^\pm$
given by $|U| = e^{-\kappa x^-}/\kappa$ and 
$|V| = e^{\kappa x^+}/\kappa$. 

The Hartle-Hawking state represents a state in equilibrium 
with the black hole and is uniquely characterized by its
global nonsingularity and its isometry invariance under the Killing time
\cite{YKHH}.
We substitute $\hOmega=\Omega_{HH}$ in Eq.(\ref{eq:CHANGEE}) 
and use $(U,V)$ for $(\hU,\hV)$, respectively.
Using the relation $dV_0=\exp(\kappa x^+_0) dx^+_0$
and the fact that ${\Hp}$ is a Killing horizon ($\partial_+\kappa=0$),
we obtain the change in energy for the Hartle-Hawking state as
\begin{equation}
	{d E_{HH}(x^+_0) \over d x^+_0} = -T_{++}|_{\Hp} = 0~.
\end{equation}
This result is expected one from the fact that 
the energy flow lines of the Hartle-Hawking state are 
along the orbits of the Killing vector
because the Hartle-Hawking state is stationary state 
with respect to the Killing time.

On the other hand, the Unruh state represents, on 
the eternal black hole, a state which is 
in a gravitationally collapsing spacetime.
Substituting $\hOmega=\Omega_{U}$, and
using $(U,x^+)$ for $(\hU,\hV)$, respectively, we get
$T_{++}|_{\Hp}=-\kappa^2/(48\pi)$ 
at the future event horizon, and obtain the result
\begin{equation}
	{d E_U(x^+_0) \over d x^+_0} = +{\kappa^2 \over 48\pi} 
	= +{\pi \over 12} (\TBH)^2~,
\label{eq:APPE}	
\end{equation}
where we used the fact that the Hawking temperature 
is given by $\TBH={\kappa \over 2\pi}$.
This represents the energy density of a right-moving 
one dimensional massless gas with the temperature $\TBH$ 
and can be interpreted as the outgoing energy flux 
due to the Hawking radiation 
(Appendix~\ref{sec:append}).


\section{The calculation of the entanglement entropy}

\label{sec:evolII}


In this section, we will calculate the entanglement entropy  
$S_M$ for the state of the fields outside the event horizon.
The concept of the quantum entanglement entropy is associated with
the notion of coarse graining associated with a division of 
the Hilbert space of a composite system.
The division may be introduced by dividing 
the whole degrees of freedom into accessible(system in interest)
and inaccessible ones(environment).
For instance, in a spacetime with black holes,
it is natural to take degrees of freedom outside
the event horizons as the accessible ones.
The density matrix appropriate to the system in interest
is obtained by tracing the whole density matrix $\hrho_{whole}$
over the environment
\begin{equation}
	\hrho_{sys} = Tr_{env} \left( \hrho_{whole} \right)~.
\label{eq:ENT1}
\end{equation}
This reduced density matrix no longer describes a pure state 
generally, even though the whole system is pure.

And then, we define the entropy of the system in interest as
\begin{equation}
	S_{sys} = -Tr_{sys} [\hrho_{sys} \ln \hrho_{sys}]~.
\label{eq:ENT2}
\end{equation}
The quantity $S_{sys}$ describes correlations between 
the system in interest and the environment, and measures the information 
which is lost by tracing over the environment. 
Note that when the whole system is in a pure state,
there is a symmetry with 
respect to an exchange of the system in interest for the environment;
the two density matrices obtained by tracing over 
accessible degrees of freedom or inaccessible ones 
give the same entropy
\cite{MKSK}.
When the whole system is in a mixed state, 
this symmetry hold no longer generally.

We are interested in the entanglement entropy of a local quantum field
associated with the division of degrees of freedom by partitioning
a time slice $\Sigma$ into accessible region $D$ and inaccessible one
$\Sigma - D$.
Then, the entanglement entropy is invariant for local deformations of the
time slice which keep the boundary $\partial D$ fixed in the spacetime.
This fact follows the unitary evolution of the whole system 
and the local causality
because, in this case, the unitary evolution operator for the whole
becomes a product of two commutable unitary operators,
each of which is the evolution operator associated with the deformation
of the time slice in the accessible or the inaccessible region.
In this sense, the entanglement entropy of any local field 
is a quantity connected with the boundary of the accessible region.


\subsection{Entanglement entropy in flat spacetime}


First, we consider the Minkowski vacuum of
a massless conformal scalar field in flat two dimensional spacetime
and calculate the entanglement entropy.

The metric of the flat spacetime is written by
\begin{equation}
	ds^2 = -dU dV~.
\end{equation}

We want to compute the entanglement entropy of the Minkowski vacuum
when the accessible region is given by the interval between 
the point (inner boundary) $P_0=(U_0,~V_0)$ 
and the point (outer boundary) $P_1=(U_1,~V_1)$.

To proceed this calculation, we need to expand 
the Minkowski vacuum by the states which live in the Hilbert space
associated with the accessible region and the ones which live in 
the Hilbert space associated with the inaccessible region.
Such decomposition can be achieved by introducing the Rindler chart
such that the point $P_0$ corresponds to the bifurcate point.
This is given by Fiola, et.al.
\cite{FIOLA},
following Unruh's calculation
\cite{UNRUH}.

Then, we can derive the entanglement entropy by standard procedure
using Eq.(\ref{eq:ENT1}) and Eq.(\ref{eq:ENT2}).
The result is given by
\cite{FIOLA}
\begin{equation}
	S={1 \over 6} \left[ \left( \ln D - \ln \epsilon_0 \right)
	+ \left( \ln D - \ln \epsilon_1 \right) \right]~,
\label{eq:FLAT}	
\end{equation}
where $D$, $\epsilon_0$ and $\epsilon_1$ are the size of 
the accessible region defined by $D^2=|(V_1-V_0)(U_1-U_0)|$,
the short distance cutoff at $P_0$ and $P_1$, respectively.

Next, we examine the change of entropy between 
two adjacent time slices.
We consider the case that the outer boundary $P_1$ is fixed
and the inner boundary $P_0$ moves along the null line 
$U_0=\mbox{constant}$.
Then, we get
\begin{eqnarray}
	{d S \over dV_0}\Big{|}_{U_0=const.}=
	-{1 \over 6}{1 \over V_1-V_0} <0 ~,
\end{eqnarray}
where the proper lengths $\epsilon_0$ and $\epsilon_1$ are fixed.

This decrease of the entanglement entropy can be understood 
as a result of the decrease of the size of the accessible region
(see Sec.\ref{sec:phys} for the more intuitive explanation of 
this result.)


\subsection{Entanglement entropy in curved spacetime}


In this subsection, we will generalize the result 
in the previous subsection to the case of curved spacetime.

Since any two dimensional spacetime is conformally flat,
we can write the metric as
\begin{equation}
         ds^2 = -{\hOmega}^2 d \hU d \hV~=-{\hOmega}^2 d{\hat{s}}^2.
\end{equation}

When we take conformal vacuum associated with the $\hU$ and $\hV$, 
as the quantum state, the expression Eq.(\ref{eq:FLAT})
can be used as it is, because the spacetime with the metric $d{\hat{s}}^2$
is flat.
Assuming that the accessible region is the interval between 
$\hP_0=(\hU_0,~\hV_0)$ and $\hP_1=(\hU_1,~\hV_1)$,
the entanglement entropy is given by
\begin{eqnarray}
	S_{\hOmega}&=&{1 \over 6} \ln 
	{ \left|\left( \hV_1 -\hV_0 \right) \left( \hU_1 -\hU_0 \right) \right|
	    \over \heps_0 \heps_1 }
\label{eq:CONENT} \\
	&=&{1 \over 6} \ln {\hOmega}_0 + {1 \over 6} \ln {\hOmega}_1
	+ {1 \over 6} \ln 
	\left| \left( \hV_1 -\hV_0 \right) \left( \hU_1 -\hU_0 \right) \right|
	- {1 \over 3} \ln \epsilon~,
\label{eq:INVFORM}	
\end{eqnarray}
where in the second line, 
we rewrite the short distance cutoffs $\heps_0$ and $\heps_1$
in the unphysical spacetime with the metric $d{\hat{s}}^2$ 
in terms of proper lengths, 
that is $\epsilon_i=\hOmega_i~\heps_i$ ($i=0,~1$) and 
set $\epsilon_0 = \epsilon_1 = \epsilon$.

Thus, the change in entropy as we move the inner boundary $\hP_0$ 
along the future horizon $\Hp$ with the proper length 
$\epsilon$ and the outer boundary $\hP_1$ fixed is given by
\begin{equation}
	{d S_{\hOmega} \over dx^+_0}={1 \over 6}
	{\partial_{+}\hat{\Omega} \over {\hat{\Omega}}}\Big{|}_{{\cal H}^+}
	-{1 \over 6}{\partial_+ \hat{V}_0 \over 
	 \hV_1 -\hV_0 }.
\label{eq:BIBUN}
\end{equation}
We apply the above result Eq.(\ref{eq:BIBUN}) to the Hartle-Hawking state 
and the Unruh one.

For the Hartle-Hawking state, we substitute 
$\hOmega=\Omega_{HH}$ in Eq.(\ref{eq:BIBUN}) and use the relation
$dV_0=\kappa V_0 dx^+_0$,
we obtain the result
\begin{equation}
	{d S_{HH} \over dx^+_0}=-{1 \over 6}
	{\kappa V_0 \over V_1 -V_0 }<0~.
\label{eq:HHH}	
\end{equation}
Against our intuition, this result shows that
the entropy $S_{HH}$ decreases as time elapses,
because of the decrease of the size of the accessible region.

For the Unruh state, substituting $\hOmega=\Omega_{U}$ 
in Eq.(\ref{eq:BIBUN}) 
and using $x^+$ for $\hV$, we obtain
\begin{eqnarray}
	{d S_U \over dx^+_0}& = &
	{\kappa \over 12}-{1 \over 6}
	{1 \over x^+_1 - x^+_0}
\\
	& = & {\pi \over 6}\TBH-{1 \over 6}
	{1 \over x^+_1 - x^+_0}~.
\label{eq:UNV}	
\end{eqnarray}
As the interpretation of Eq.(\ref{eq:APPE}), the first term in 
Eq.(\ref{eq:UNV}) can be understood as the entropy 
density of right moving one dimensional massless gas 
with temperature $\TBH$ and can be interpreted as 
the entropy production rate due to the Hawking radiation 
(Appendix~\ref{sec:append}).
Since the second term gives a negative contribution, the R.H.S. of
Eq.(\ref{eq:UNV}) can not have a definite sign. In fact, this term
grows without bound as the time slice approaches to the null one.

Note that the first term in Eq.(\ref{eq:UNV}) which has a natural 
interpretation as the Hawking radiation appears by 
fixing the proper short distance cutoff $\epsilon$, 
not but fixing the cutoff $\heps$ in the unphysical spacetime.


\section{Physical interpretation of our result}\label{sec:phys}


We summarize the results for the typical two vacuum states 
in terms of the change in  free energy.
Using the relations $\bBH=2\pi/\kappa$ and 
the definition $F_{M}\equiv E_M-\bBH^{-1} S_{M}$, 
the change in free energy is given by:

\begin{equation}
	{d F_{HH} \over dx^+_0} = {d E_{HH} \over dx^+_0}
	-{\kappa \over 2\pi}{d S_{HH} \over dx^+_0} 
	= +{{\kappa}^2 \over 24\pi}{V_0 \over 
	V_1 -V_0}>0~
\label{eq:dFHH}
\end{equation}
for the Hartle-Hawking state,
\begin{equation}
	{d F_U \over dx^+_0} = {d E_U \over dx^+_0}
	-{\kappa \over 2\pi}{d S_U \over dx^+_0} 
	= -{\kappa^2 \over 48\pi}+{\kappa \over 24\pi}
	{1 \over x^+_1 -x^+_0}~
\label{eq:dFU}
\end{equation}
for the Unruh state.

Where, if we take the limit that our accessible region extends to 
the spatial infinity, i.e., $V_1,~x^+_1 \rightarrow \infty$,
we obtain the desired result for the GSL.
However, provided that we hold the accessible region finite,
the above results show that the free energy (total entropy) 
does not necessarily decrease (increase) and 
the time derivative form of the GSL does not always hold 
for two dimensional eternal black hole background.
Note that the second term in Eq.(\ref{eq:dFU}) can grow unboundedly
as time evolves.
We can recognize that the violation of the GSL is caused by the decrease 
of the entanglement entropy of fields associated with the decrease 
of the size of the accessible region.
The change in the entanglement entropy comes from two parts:
one is associated with ultraviolet divergent term which represents 
short distance correlations between the modes near the horizon; 
and the other is associated with the size of the accessible region
which contains long distance correlations between the modes 
far inside and far outside the horizon
(Hereafter, we call this term infrared divergent term.). 
The ultraviolet divergent term can be interpreted 
as the entropy production due to the Hawking radiation 
for the Unruh state
and it can be thought to give non-negative contribution 
to the change in total entropy.
On the other hand, the infrared divergent term gives 
negative contribution due to the decrease of the size
outside the event horizon and it is just this behavior that causes
the violation of the GSL.

When the outer boundary exists at a finite distance,
we do not observe the whole external region outside the horizon.
Therefore one might think that
the violation of the GSL is brought about by interaction with 
the field degrees of freedom in the rest external region 
that we do not observe.
That is, one suspect that the composite system 
composed of the black hole and the field degrees of freedom in 
the accessible region might not be isolated.
However, since we fix the outer boundary,  
there is no exchange of heat, work and so on,
by the interaction between the composite system and the rest one.
Therefore, 
the composite system is isolated in substance and
our results Eqs.(\ref{eq:dFHH}) and (\ref{eq:dFU}) point to
the violation of the GSL.

Fiola, et.al.
\cite{FIOLA}
considered the GSL for a black hole formed by gravitational collapse
using the RST model
\cite{RST}
and concluded that the GSL holds in the RST model 
under suitable conditions.
Note that our conclusion is different from theirs
in spite of the fact that the entanglement entropy of fields
are used as the quantity $S_M$ for both cases.
Of course, there are various differences between these works.
Especially, a main difference is the contribution of the third term in 
Eq.(\ref{eq:INVFORM}) 
(the last term in their Eq.(75)) to the expression $dS_M/dx^+_0$.
That is, the behavior of the infrared divergent term aforementioned:
the contribution in their case is always positive,
while one in ours is negative.

The origin which causes this difference comes from the existence of
the reflecting boundary, i.e., the difference of the boundary conditions.
In the RST model, they need to impose reflecting boundary condition
at {\lq}{\lq}central point'', beyond where the dilaton becomes an imaginary
value.
Therefore, their quantum state of the scalar field has correlation between
the right moving modes and the left moving ones.
In eternal black hole background, this corresponds to the case
in which we make initial state to have correlation 
between the modes on ${\cal H^-}$ and ones on ${\cal I^-}$, 
while the states in our investigation have no correlation between them.
This brings about the difference between
the dependence of $dS_M/dx^+_0$ on $P_0$ in each case.
(So if we impose the suitable boundary condition, we could reproduce the
result corresponding to theirs.)

Next, we will give a more intuitive explanation of this difference 
by using quantum correlation between two wave packets of matter fields.

\begin{figure}
\begin{center}
\leavevmode
\epsfysize=5.0cm
\epsfbox{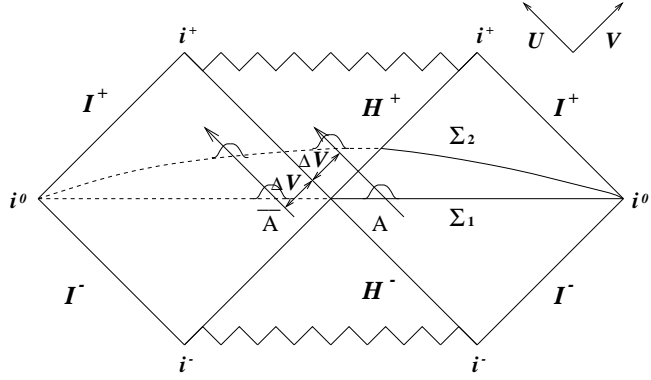}
\end{center}
\caption[]{The most correlated pair $(A,\bar{A})$ on eternal black hole 
background.
For an observer that locates at outside of the horizon, the contribution
of this pair to the entropy
decrease as time evolves and cross section $\Sigma_t \cap{\cal H^+}$ move
forward along ${\cal H^+}$.}
\label{fig:correl1}
\end{figure}

We first consider the case for an eternal black hole. 
It is one example of the spacetime without a reflecting boundary.
We take the conformal vacuum states and consider two adjacent time slices 
$\Sigma_1$ and $\Sigma_2$ and notice one of the most correlated pairs
\footnote{
See 
\cite{TAKAGI}
for the example of the Minkowski spacetime.
The same argument can be applied to the black hole
by choosing suitable basis:
we can make the most entangled pairs located
at the equal null distance from $v=v_0$ (or $V=0$)
for a black hole with a boundary (or without a boundary).

}
($A$, $\bar{A}$) of 
the left-moving modes that are specified by the equal null distance 
$\Delta V$ from $V=0$
(Fig.\ref{fig:correl1}).

While we can access one of them ($A$) on $\Sigma_1$, we can access
neither of them on $\Sigma_2$ 
due to the existence of the horizon $\cal{H}^+$. 
Thus, the contribution of this pair to the entanglement entropy exists
on $\Sigma_1$ and vanishes on $\Sigma_2$.
The same argument can also be applied to all the other 
pairs. 
Since we move the inner boundary $P_0$ of $\Sigma$ 
along the future horizon $\cal{H}^+$,
this fact reflects as an decrease of entanglement entropy as time evolves.
Note that right-moving mode does not influence to the change in entanglement
entropy.
After all, the decrease of the entanglement entropy for an eternal black hole
can also be understood
by the decrease of the number of the pairs which contributes to the 
entanglement entropy.

Subsequently, we apply the above arguments to the black hole formed 
by gravitational collapse with a reflecting boundary
\footnote{Note that we can consider the black hole formed by gravitational 
collapse without a reflecting boundary: 
The shock wave solution in CGHS model, for example.}.
Note that, different from the no boundary case,
the correlation between the right moving modes and the left moving ones
is induced by the existence of the boundary in this case.
This produces an important difference from no boundary case.

\begin{figure}
\begin{center}
\leavevmode
\epsfysize=7.0cm
\epsfbox{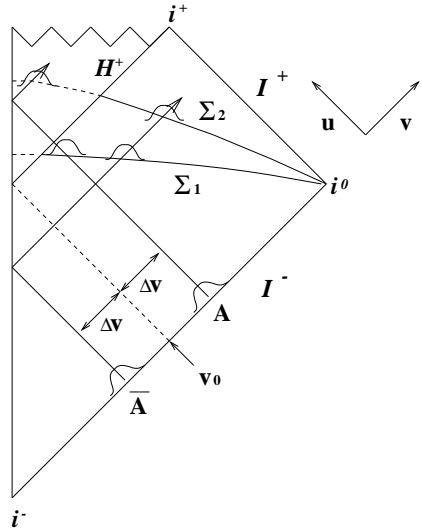}
\end{center}
\caption[]{The most correlated pair $(A,\bar{A})$ on black hole 
formed by gravitational collapse with a boundary. 
For this case, the contribution of this pair to the entropy
increases as time evolves.}
\label{fig:correl2}
\end{figure}

We introduce the null coordinates ($u$, $v$) and suppose that 
the formation of the event horizon ${\cal H}^{+}$ is at $v=v_0$.
We notice one of the most correlated pairs ($A$, $\bar{A}$) of the ingoing modes 
that are specified by the equal null distance $\Delta v$ from $v=v_0$, and
consider two adjacent time slices $\Sigma_1$ and $\Sigma_2$
(Fig.\ref{fig:correl2}).

Since we can access both modes ($A$, $\bar{A}$) on $\Sigma_1$
, the contribution of this pair to the entanglement entropy is zero. 
But on $\Sigma_2$, since we can access only one of them ($\bar{A}$), 
nonzero contribution is produced.
As the same argument can be repeated for all the other pairs, 
the entropy is increasing as time evolves in the collapsing model
with a boundary.

After all, we can understand the behavior of the infrared divergent term
intuitively:
it gives a positive contribution to the change in entropy 
in the cases with a boundary, and gives a negative contribution
in the ones without a boundary.
In other words, we can say that our work gives the situation where 
the entanglement entropy increase (or decrease) in time.


\section{Summary and Discussion}\label{sec:summary}


In this paper, we examined the validity of the GSL for 
a black hole in two dimensional gravitational theories
under the quasistationary approximation.
In order to satisfy the quasistationary approximation,
we considered the infinitesimal time development of the total entropy 
of the black hole and the field degrees of freedom outside the horizon.
Our approach can be applied to test the validity of the GSL in any
two dimensional stationary black hole spacetime
which possesses the first law of black hole mechanics, 
irrespective of models.

Making use of the fact that the change in total entropy
is equal to minus the change in free energy of the fields
outside the horizon under the quasistationary approximation,
we calculated the change in free energy for
the conformal vacuum states. 
In particular, we applied the result to the Hartle-Hawking state 
and the Unruh one in the eternal black hole background.
And then, we showed the differential form of the GSL to be invalid 
when our accessible region is finite and to be valid 
for infinite accessible region.

We recognized that the origin of the violation of the GSL is
the decrease of the entanglement entropy of fields associated with
the decrease of the size of the accessible region.
However, the behavior of this term is something curious 
in our intuitive terms.
Because it is usually thought that the total entropy does not change for
the Hartle-Hawking state in eternal black hole background, 
which describes the thermal equilibrium state
between the black hole and the surrounding matter field.
So, it is natural to expect that the entropy production for
the Hartle-Hawking state does not occur.

Myers
\cite{MYERS}
and
Hirata, et.al.
\cite{MHFS}
examined the validity of the GSL by using the Noether charge method
and taking into account 1-loop quantum back-reaction and 
showed that it is satisfied for both the RST model 
and the wide class of the CGHS model. 
In their analysis, the third term in Eq.(\ref{eq:INVFORM}) does not appear.
This might suggest that the infrared divergent term 
could be dropped by some physical reasons,
though, further argument about excluding this term is necessary
\footnote{
Since we consider the massless field, the term proportional to
$\ln{\hat{D}}$ appears in the result
(\ref{eq:INVFORM}).
However, if we consider the massive field, 
this term would not appear
(inverse of the fixed mass $1/m$ of the field enters into the expression 
instead of $\hat{D}$).

}. 


Anyway, it is necessary to subtract this term and 
define the total entropy to rescue the GSL.
Therefore, from now on, we suppose that this subtraction is performed 
systematically, and argue the GSL for this new quantity.
That is, we define the new entropy by
\begin{eqnarray}
	 S^{\ast}_{\hat{\Omega}} &=& 
	S_{\hat{\Omega}}
 	- {1 \over 6}\ln \left({\hat{D} \over \epsilon}\right),
\end{eqnarray}
where $\hat{D}$ is the size of the accessible region 
in the unphysical spacetime in which the conformal vacuum is defined
and $\epsilon$ is a proper length of short distance cutoff.
Then,
\begin{equation}
	{d S^{\ast}_{\hat{\Omega}} \over dx^+_0}={1 \over 6}
	{\partial_{+}\hat{\Omega} \over {\hat{\Omega}}}\Big{|}_{{\cal H}^+},
\end{equation}
and we obtain the final result
\begin{eqnarray}
	{d F^{\ast}_{\hat{\Omega}} \over dx^+_0} &=& -{\lambda \over 2\pi}
	{d S^{\ast}_{\hat{\Omega}} \over dx^+_0}+{d E_{\hat{\Omega}} \over dx^+_0}
\\
	&=& -{1 \over 12\pi}{ {\partial^2_{+}{\hat{\Omega}}} \over {\hat{\Omega}} }\Big{|}_{{\cal H}^+}
\end{eqnarray}
Therefore, the validity of the GSL depends 
only on the signature of the 
${\partial^2_{+}{\hat{\Omega}}}$ at the future horizon ${\cal H}^+$.

If we apply this result to the typical two states aforementioned,
we obtain the desired result: $dF^{\ast}_{HH}=0$ 
for the Hartle-Hawking state and $dF^{\ast}_{U}<0$ for 
the Unruh state.
However, 
since we can choose the vacuum state as we like
(in other words, we can perform conformal transformation freely), 
we can violate the GSL by choosing the suitable conformal vacuum 
which satisfy ${\partial^2_{+}{\hat{\Omega}}} |_{{\cal H}^+}<0$
even if we can subtract the infrared divergent term 
by some physical procedure successfully.

Now we consider whether the violation of the GSL occurs or not
for the wide range in the following sense:
the state that violate the GSL is not a special one;
and the violation occurs sufficiently long time.
In our case, there seem to exist a lot of states which satisfy 
both of these conditions.
Of course, since the vacuum states that we prepare should be a physically
reasonable ones, there would be some requirement from physical nature.
For example, the expectation value of the energy-momentum tensor 
should be finite at ${\cal H}^+$. But, in practice, this condition can not 
impose any condition on 
${\partial^2_{+}{\hat{\Omega}}}|_{{\cal H}^+}$,
and we can not remove the cases that violate the GSL by this criterion.
Further, noting that since ${\hat{\Omega}}$ is a function of 
the spacetime point,
it is possible to violate the GSL during sufficiently long time interval, 
by choosing a suitable form of the function.

One of such examples is the case 
${\hat{\Omega}}^2= \xi^2 /{\left[ \kappa |U|\cosh{(\kappa x^+)}\right]}$.
In this case, the behavior of the energy momentum tensor is regular
at ${\cal H}^+$ and ${\cal I}^{\pm}$,
so it seems to be a physically reasonable state. 
And then, it approaches the Hartle-Hawking
state asymptotically at ${\cal I}^+$~(as $x^+\rightarrow \infty$) 
while keeping ${\partial^2_{+}{\hat{\Omega}}}|_{{\cal H}^+}<0$.
Thus, the time duration during which the violation of the GSL
occurs can be taken arbitrary long.
It would not seem that this is a special case, because 
we can find a lot of examples which give a similar behavior as this one.

Therefore, it seems that the violation of the GSL occurs for a rather
wide range of vacuum states unless there exists a {\lq}{\lq}selection rule''
that all physically acceptable states should satisfy 
the condition ${\partial^2_{+}{\hat{\Omega}}}|_{{\cal H}^+}>0$.
Then, considering that our analysis is independent of models,
we would have only two choice to resolve the violation of the GSL.
One is that the states which satisfy 
${\partial^2_{+}{\hat{\Omega}}}|_{{\cal H}^+}<0$ like an above example
should be excluded by some physical requirement that we can not find now. 
It means that all the physically reasonable states are restricted to ones 
that satisfy the condition ${\partial^2_{+}{\hat{\Omega}}}|_{{\cal H}^+}>0$.
The other is that we improve the entropy formula $S^{\ast}$ further.
That is, the violation of the GSL was thought to be caused 
by the wrong definition of the entropy.

In either cases, further investigation would be necessary and
resolution of the difficulty must be brought over the future works.

\acknowledgments
The authors would like to thank M. Shibao
for valuable comments and stimulating discussions.
We also appreciate Professor A. Hosoya for continuous encouragement.
\appendix
\section{}\label{sec:append}

We can interpret the results, Eqs.(\ref{eq:APPE}) and 
(\ref{eq:UNV}) as the energy and entropy production rate 
due to the Hawking radiation. 

An observer at the null infinity observes quanta distributed 
per mode and per unit time by
\begin{equation}
	<n_\omega> = {\Gamma_\omega \over 
	\exp(\omega/~\TBH)-1 }~,
\end{equation}
in a two dimensional black hole background 
with the temperature $\TBH$.
The graybody factor for an massless conformal scalar field is
$\Gamma_\omega=1$ because of no scattering.

Therefore, the energy production rate is given by
\begin{equation}
	{dE_{rad} \over dt} = {1 \over 2\pi} \int^\infty_0 d\omega~ 
	\omega {\Gamma_\omega \over \exp(\omega/~\TBH)-1 }~.
\end{equation}
Assuming the canonical distribution, 
the entropy per mode is given by
\begin{equation}
	S_\omega = (1+ <n_\omega>)\ln (1+<n_\omega>)
	-<n_\omega> \ln <n_\omega>~,
\label{eq:ent}
\end{equation}
and then, the entropy production rate of the emitted radiation 
is given by
\begin{equation}
	{d S_{rad} \over dt} = {1 \over 2\pi} \int^\infty_0 d\omega~
	\left[ (1+ <n_\omega>)\ln (1+<n_\omega>)
	-<n_\omega> \ln <n_\omega> \right]~.
\end{equation}

The above quantities become for a massless conformal scalar field
($\Gamma_\omega=1$) as
\begin{eqnarray}
	{dE_{rad} \over dt} &=& 
	{\TBH^2 \over 2\pi}\int_0^{\infty} dx~ {x \over e^x -1 } 
	={\pi \over 12} \TBH^2~,
\\
	{dS_{rad} \over dt} &=& 
	{\TBH \over 2\pi}\int_0^{\infty} dx~ 
	\left[ {x \over e^x -1 } - \ln (1-e^{-x}) \right]
	={\pi \over 6} \TBH~.
\end{eqnarray}
These coincide with Eq.(\ref{eq:APPE}) and 
the first term in Eq.(\ref{eq:UNV}).


\end{document}